\documentstyle[12pt,moriond,psfig]{article}
\def\lfir{$L_{\rm FIR}$}

\def\etal{et al.}
\def\hii{H{\sc ii}}

\def\micron{$\mu$m}

\def\msun{\ifmmode {M_\odot} \else M$_{\odot}$\fi}
\def\zsun{Z$_{\odot}$}
\def\halpha{\ifmmode {\rm H{\alpha}} \else $\rm H{\alpha}$\fi}
\def\hbeta{\ifmmode {\rm H{\beta}} \else $\rm H{\beta}$\fi}
\def\msunyr{\ifmmode {M_\odot \, {\rm yr}^{-1}} \else M$_{\odot}$ yr$^{-1}$\fi}
\def\mup{\ifmmode {M_{\rm up}} \else $M_{\rm up}$\fi}
\def\mlow{\ifmmode {M_{\rm low}} \else $M_{\rm low}$\fi}
\def\luv{\ifmmode {L_{\rm UV}} \else $L_{\rm UV}$\fi}
\def\lfir{\ifmmode {L_{\rm FIR}} \else $L_{\rm FIR}$\fi}
\def\lbol{\ifmmode {L_{\rm Bol}} \else $L_{\rm Bol}$\fi}
\def\l1500{\ifmmode {L_{1500}} \else $L_{1500}$\fi}
\def\lhalpha{\ifmmode {L_{\rm H\alpha}} \else $L_{\rm H\alpha}$\fi}
\def\loii{\ifmmode {L_{\rm OII}} \else $L_{\rm OII}$\fi}
\def\oii{[O~{\sc ii}] $\lambda$3727}
\def\Oii{[O~{\sc ii}]}

\begin{document}
\heading{DETERMINING STAR FORMATIONS RATES: METHODS AND UNCERTAINTIES
}

\author{D. Schaerer $^{1}$}{$^{1}$ Observatoire Midi-Pyr\'en\'ees, F-31400 Toulouse, France.
  (schaerer@obs-mip.fr)}{}{}

\begin{moriondabstract}
Methods for the determination of star formation rates (SFR) from integrated populations
are reviewed. We discuss the assumptions and underlying uncertainties (e.g.\ IMF slope, 
\mup, metallicity, SF history etc.) used in the calibrations of UV, FIR, \halpha, and 
\oii\ indicators.
The ``universality'' of empirically calibrated indicators such as \Oii\ is examined.
We also present preliminary results from a theoretical study with the aim to
understand the systematics of forbidden line indicators and to provide new SFR indicators
including optical to far-IR lines.

\end{moriondabstract}

\section{Introduction}
Determinations of the past and current rate of star formation are important for the
understanding of a wide variety of astrophysical problems ranging for example
from studies of physical processes in ``local'' objects,
over the origin of the Hubble sequence of galaxies, to 
different scenarios for the formation of galaxies and cosmological structures.
Correspondingly the observables used to determine a star formation rate (SFR) vary 
greatly from local measures (e.g.\ color-magnitude diagrams of resolved populations), 
over integrated spectra of galaxies, to an average luminosity density
representative of the cosmic history of star formation. 
In the present review we cover only methods based on the analysis of light from
{\em integrated populations}, which are of interest for the study of distant objects.
Many interesting results have also been obtained in the recent years from 
analysis of the stellar content and the star formation history in {\em resolved objects}.
For an overview of this subjects and references to different techniques 
the reader is e.g.\ referred to the review of Mateo (1998). 

Essentially four basic and widely used methods can be identified serving determinations of
the star formation rate from integrated observations
(see also review of Kennicutt 1998a): 
1) UV continuum methods, 2) Far-IR and radio continuum methods, 3) analysis based on 
recombination lines, and 4) forbidden lines.
These will be discussed individually below.
In general, these direct methods, can be applied to individual star forming regions,
and observations of individual galaxies, as well as to populations of galaxies.
For the latter, and in particular for studies aiming to derive the global
star formation history of the universe, obviously numerous additional
constraints exist (e.g.\ luminosity functions, cosmic background, SN rates, $\gamma$-ray
bursts etc.). Since amply discussed in other contributions to this conference,
these more indirect techniques, which generally also depend on additional
parameters, will not be discussed here.

The outline of this review is as follows.
The general procedure and basic assumptions common to all four methods are
summarized in Sect.\ \ref{s_gen}.
These methods are then discussed individually in Sects.\ \ref{s_uv} to \ref{s_oii}.
New results from a theoretical study of various line indicators are
presented in Sect.\ \ref{s_new}.
A summary and conclusions are presented in Sect.\ \ref{s_con}.

\section{General assumptions and procedure}
\label{s_gen}

All SFR determinations rely on a calibration relating the energy output in the
considered wavelength range to the total stellar mass.
This is usually done using predictions from an evolutionary synthesis model.
The basic input parameters of these models are: 
1) the metallicity of the stars, 2) the star formation history, 
3) a description of the IMF, 4) stellar tracks, and 5) stellar atmospheres.
%
These input parameters and all related uncertainties affect directly all SFR 
calibrations. 

To which degree the SFR calibrations depend on parameters 3-5 will
be illustrated by performing test calculations using sets of standard and 
``alternate'' input parameters. 
For this purpose we have used the synthesis models of Schaerer \& Vacca
(1998) and the recent {\em Starburst99} models of Leitherer \etal\ (1999).
The results are shown in the subsequent sections.
The complete set of parameters used is given in Table \ref{ta_pars}.
Since all observables used here are only sensitive to stars with masses
$\gsim$ 2-5 \msun, the lower mass cut-off \mlow\ affects only the
absolute normalisation of the SFR (see below).
\mup=30 \msun\ is used to examine cases with a lack of massive stars, as suggested
for some cases in the literature (e.g.\ Rieke \etal\ 1980, Goldader \etal\ 1997).
$\alpha=2.7$ is the slope of the Scalo (1986) IMF derived for $M > 2 \msun$.
It allows us to study the effect of an IMF steeper than Salpeter.
 
For sound comparisons between different SF indicators it is necessary
to make sure that the assumptions made for their calibration are consistent.
This requires in particular the use of the same IMF, which is not always
the case in the literature. To facilitate this task we adopt 
a Salpeter IMF from 0.1 to 100 \msun\ for convenience and for direct
comparison with the review of Kennicutt (1998a, hereafter K98).
However, it is important to note that SFRs derived with this assumption
are {\em overestimated} typically by a factor of 2.6--5.5 compared to 
SFR determinations taking the flattening of the IMF below 1 \msun\ into account
\footnote{Over the mass interval 0.1--100 \msun\ the Salpeter IMF yields a
total mass larger by a factor 2.6 compared to the Kroupa \etal\ (1993) IMF,
2.9 compared to Kroupa (1998), 2.8 compared to Kroupa (1998) with a Salpeter slope
above 1 \msun, 5.5 compared to Reid \& Gizis (1997), and a factor 4.4-4.7
compared to Scalo (1986 and 1998).
A single powerlaw IMF scales as $M/M^\prime=(\mup^{2-\alpha}-\mlow^{2-\alpha}) /
({M^\prime_{\rm up}}^{2-\alpha}-{M^\prime_{\rm low}}^{2-\alpha})$, where
$\alpha=x+1$ is the IMF slope. E.g.\ for the Salpeter IMF ($\alpha=2.35$)
a change of the lower mass cut-off 
from 0.1 to 1 \msun\ corresponds to a decrease of the total mass by a factor
2.56 for \mup=100 \msun.}.
For detailed discussions on the IMF the reader is referred to the recent 
conference volume by Gilmore \etal\ (1998).

Before theoretical SFR calibrations can be used the observational
data must of course be corrected for extinction. The importance of this 
``correction'', especially for UV observations, is well recognised and
many different procedures have been established (cf.\ below; see also Calzetti, 
these proceedings).

\begin{table}
\caption{Input parameters of synthesis models used to illustrate dependence
of SFR calibrations on the IMF (slope, \mup) and metallicity.}
\label{ta_pars}
\begin{tabular}[htb]{lllll}
IMF & \mlow\ & \mup\ & Metallicity & Symbol/color \\
\hline
Salpeter ($\alpha=2.35$) & 0.1 & 100 & solar & solid/black (``standard model'')\\
idem  &  0.1 & 100 & 1/20 \zsun\ & dotted/black \\
idem  &  0.1 & 100 & 2 \zsun\    & long-dashed/black \\
idem  &  0.1 & 30  & solar       & dash-dotted/green \\
$\alpha=2.35$ ($M \le 1 \msun$) & \\
$\alpha=2.7$ ($M > 1 \msun$)&  0.1 & 100  & solar       & short-dashed/red \\
\hline 
\end{tabular}
\end{table}

\begin{figure}[tb]        
\centerline{
\psfig{figure=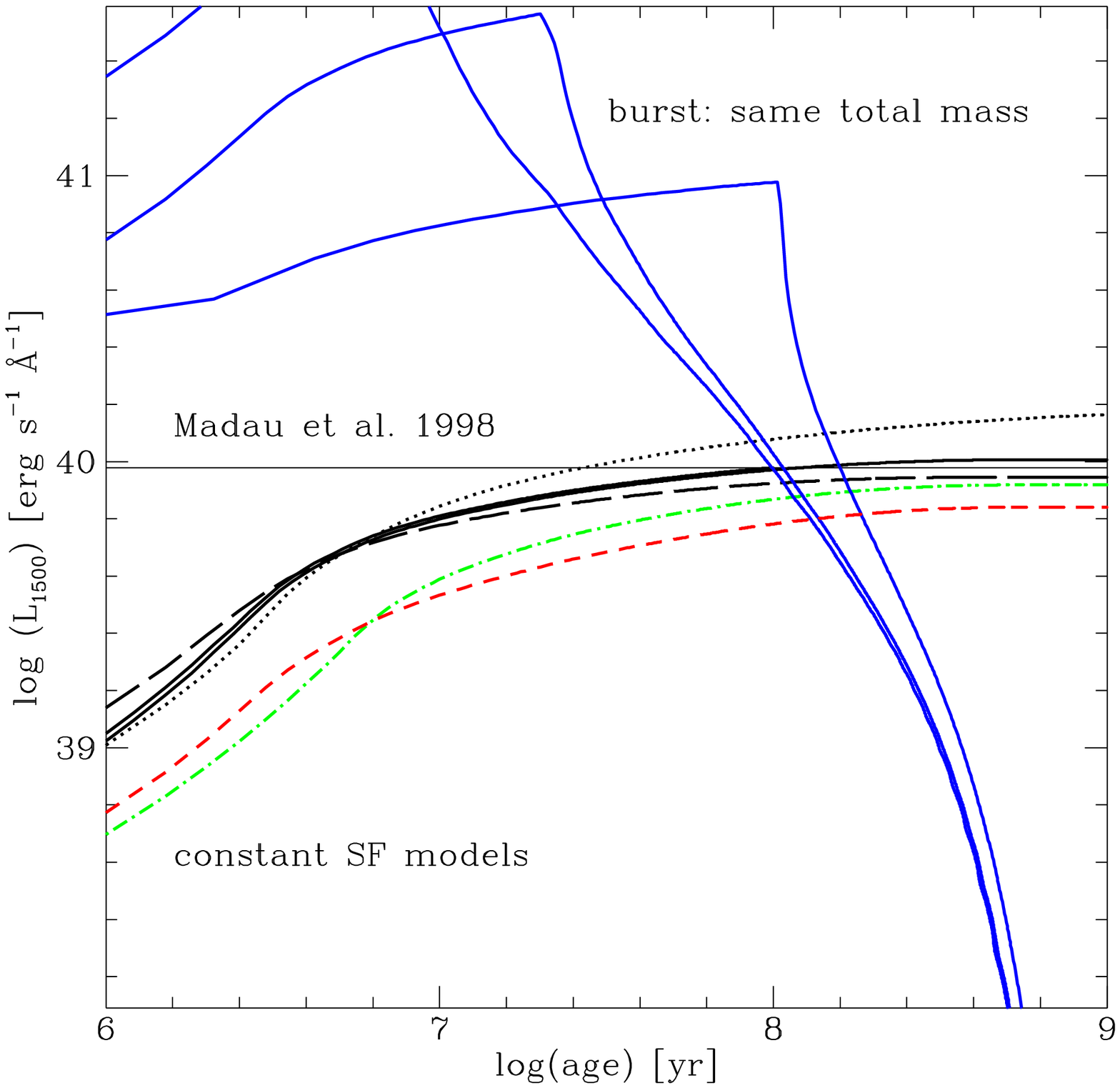,width=8cm}
\psfig{figure=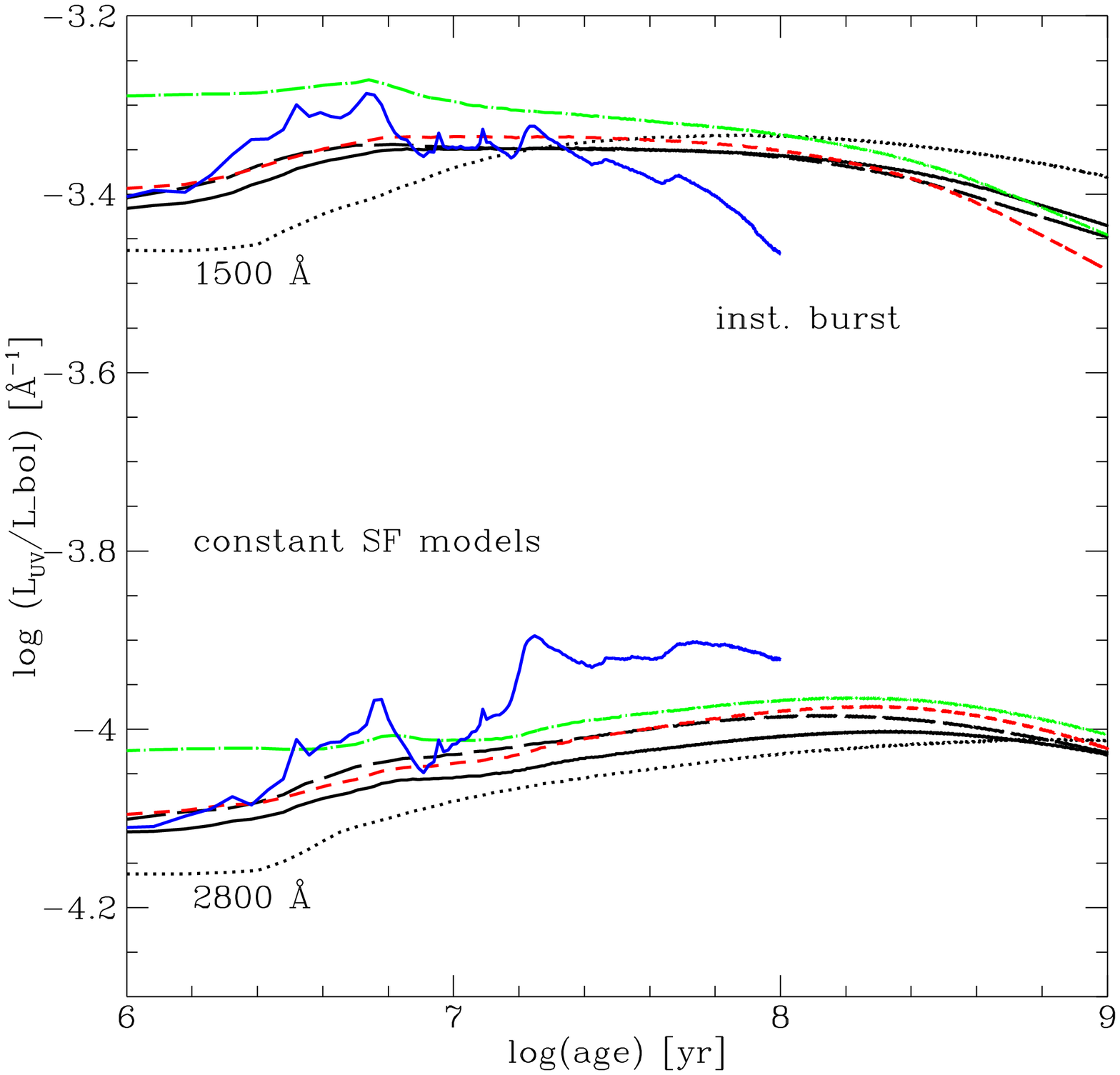,width=8cm}}
\caption{{\em Left:} Temporal evolution of \l1500\ for models with constant SF
($SFR=1$ \msunyr, see Table \ref{ta_pars} for symbols), and burst models.
The SFR(UV) calibration from Madau \etal\ (1998, our Eq.\ \ref{eq_uv}) is shown
by the horizontal line.
{\em Right:} \luv/\lbol\ for 1500 and 2800 \AA\ for const SF and instantaneous
burst models.}
\label{fig_uv1}
\end{figure}

\section{SFR from the UV continuum}
\label{s_uv}
The UV continuum generally probes emission from young stars so that it is a 
reasonable measure of ongoing star formation.
The exception are elliptical galaxies and spiral bulges showing the ``UV upturn'' 
phenomenon, likely due to so-called AGB-manqu\'e stars (e.g.\ Dorman \etal\ 1995).
The optimal wavelength range is $\sim$ 1250-2500 \AA, longward of the Ly$\alpha$
forest but at wavelengths short enough to minimize the contribution from older
stellar populations. 
References to UV observations are e.g.\ found in the review of K98.

Many different calibrations of the SFR from the UV flux have been published
(e.g.\ Buat \etal\ 1989, Deharveng \etal\ 1994, Meurer \etal\ 1995, Cowie \etal\ 1997,
Madau \etal\ 1998) for $\lambda \sim$ 1500-2800 \AA.
According to K98 these calibrations differ by up to $\sim$ 0.3 dex when converted
to a common wavelength and IMF, and differences stem from the use of various 
stellar libraries and different assumptions of the star formation timescales.
The latter effect seems to be dominant (cf.\ below).


For a Salpeter IMF from 0.1 to 100 \msun\ the calibration of Madau \etal\ (1998)
yields
\begin{equation}
  \label{eq_uv}
  {\rm SFR} (\msun \, {\rm yr}^{-1}) = 1.4 \times 10^{-28} L_\nu 
  ({\rm erg \, s^{-1} \,Hz^{-1}}) 
\end{equation}
according to K98. This expression is valid for wavelengths $\sim$ 1500-2800 \AA, since
the resulting UV spectrum is nearly flat in $L_\nu$ (cf.\ K98).
It has been obtained from the latest Bruzual \& Charlot (1998) models and represents the 
asymptotic UV flux obtained for a exponentially decreasing SFR (cf.\ Madau \etal\ 1998).

To illustrate the dependence on metallicity, IMF, and star formation history
we show in Fig.\ \ref{fig_uv1} (left) the temporal evolution of the UV luminosity 
$L_{1500}$ obtained from different models (see Table \ref{ta_pars} for parameters
and symbols used). 
After an initial increase the UV luminosity of the constant SF model reaches an 
asymptotic value at ages $\gsim$ 10$^{8-9}$ yr, which is essentially 
identical to the one from Eq. \ref{eq_uv} obtained for SFR $\propto \exp^{-t/\tau}$. 
This relatively long ``equilibrium timescale'' ($\tau_{\rm UV} \sim$ 1 Gyr) implies
in particular that at very high redshift (typically $z \gsim $4) this situation may 
not be reached yet.
For younger bursts producing less UV light for a given star formation rate, the SFR 
derived from \luv\ would be higher than given by Eq.\ \ref{eq_uv}.
The calculations for metallicities between 1/20 and 2 \zsun\ show typically differences
of 0.2 dex.
Decreasing the upper mass cut-off to 30 \msun\ or using a steeper
IMF slope changes the \luv--SFR relation by a similar amount.
We identify the spread between all constant SF models at equilibrium ($\sim$ 1 Gyr)
as a ``typical uncertainty'' due to the metallicity dependence, and the slope and
\mup\ of the IMF. For SFR(\luv) this is found to be $\sim$ 0.4 dex.

In Fig.\ \ref{fig_uv1} (left) we also show three burst models with a duration of 5, 20, and
100 Myr forming the same mass of stars (i.e.\ $10^9$ \msun) as the constant SF models 
over 1 Gyr. Obviously such scenarios show a very different temporal evolution of the 
UV light, and no simple \luv-SFR relation holds.
However, if the observations include a sufficiently large number of such SF regions
sampling all ages, Eq.\ \ref{eq_uv} yields again the correct total SFR.

The right panel in Fig.\ \ref{fig_uv1} shows the ratio of \luv/\lbol\ for instantaneous
(delta) bursts and constant SF. A fairly tight relation is expected between these
quantities, quite independently of the SF history. This property may be used for 
SFR cross calibrations between the IR and UV (see Sect.\ \ref{s_ir}).

As mentioned earlier, extinction is a major issue regarding the determination
of SFRs from the UV. Although justice cannot be given to this subject here,
it is, however, useful to recall 
a particular feature of the UV spectrum which enables such corrections in an
efficient way. Indeed, a fairly narrow range of values of the UV slope $\beta$ is 
expected, at least for constant SF (e.g.\ Meurer \etal\ 1995, 1997). 
Together with the finding of empirical correlations between $\beta$ and the extinction 
from the Balmer decrement (Calzetti \etal\ 1994, 1996) and $\log(F_{\rm FIR}/F_{2200})$ 
which shows that dust absorption correlates with UV reddening (Meurer \etal\ 1995, 1997), 
this can be used to correct, at least in a statistical sense, for extinction.
This method has been successfully applied to starburst galaxies by Meurer \etal\
(1997, 1999). Other methods have e.g.\ been used by Buat and coworkers (e.g.\ Buat 1992,
Buat \& Xu 1996, Buat \& Burgarella 1998).
Additional information and a summary of studies related to star formation rates using 
UV methods is found in K98.

{\em Summary:}
From the above we conclude that the most important assumptions entering the
calibration of SFR from the UV continuum are: the SF history, the IMF slope and
\mup, the latter two being approximately of the same importance.
The ``typical uncertainty'' (as ``defined'' by the test calculations shown above)
due to metallicity, IMF slope, and \mup\  is $\sim$ 0.4 dex.
Extinction corrections, treated as ``external'' in this context, are obviously of 
prime importance for this SFR indicator.

\section{SFR from the far-IR and radio continuum}
\label{s_ir}
\subsection{Far-IR methods}
At the basis for the use of the far-IR (FIR) continuum as a measure of star formation
are the facts that 1) a significant fraction of the bolometric luminosity of a galaxy
is absorbed by interstellar dust and re-emitted in the thermal IR, and 
2) the absorption cross section of dust strongly peaks in the UV which
traces young stellar populations.

The most simple assumption used for calibrations with synthesis models is that the 
FIR luminosity (as e.g.\ measured from 8-1000 \micron)
represents the {\em total bolometric} luminosity, or in other words
that the dust is optically thick in the SF regions. This may e.g.\
be the case in IR luminous galaxies. In general the physical situation is, however, 
more complex. Different dust components (warm dust, cirrus) can be found, old stars
and AGN may contribute to the heating of dust, and the optical thickness (or equivalently
the reprocessing or transfer efficiency) may vary (see e.g.\ references in K98).
No accurate universal SFR(IR) indicator can thus be expected.

\begin{figure}[tb]        
\centerline{
\psfig{figure=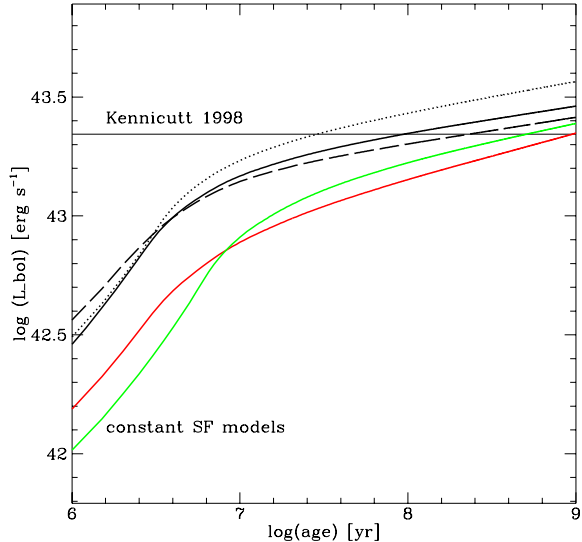,width=8cm}}
\caption{Same as Fig.\ \ref{fig_uv1} (left) for models with $SFR=1$ \msunyr.
The Kennicutt (1998a) relation is shown by the horizontal line.}
\label{fig_fir}
\end{figure}

Various calibrations based on synthesis models have been published 
(e.g.\ Hunter \etal\ 1986, Meurer \etal\ 1997, Kennicutt 1998b).
With the assumption \lfir=\lbol\ and using the same IMF as in Eq.\ \ref{eq_uv}
\begin{equation}
  \label{eq_fir}
  {\rm SFR} (\msun \, {\rm yr}^{-1}) = 4.5 \times 10^{-44} L_{\rm FIR}
  ({\rm erg \, s^{-1}}) 
\end{equation}
is obtained by K98 from the mean \lbol\ for 10-100 Myr continous bursts 
at solar metallicity according to the Leitherer \& Heckman (1995) models.
Although according to K98 most other published calibrations lie within $\pm$ 30 \%
of this equation, e.g.\ the frequently used value of Thronson \& Telesco (1986) is
a factor of 3.8 larger for the same IMF.
Most important, and larger than this, is in any case the uncertainty due to the adoption 
of an appropriate age for the stellar population.
This is also illustrated in Fig.\ \ref{fig_fir} which shows the temporal evolution
of \lbol\ for different models of constant SF.
For obvious reasons the bolometric luminosity evolves on longer timescales than e.g.\
\luv, and an ``equilibrium value'' is not reached before $\gsim 10^{10}$ yr.
An assumption on the typical age or an appropriate mean age has thus to be adopted.
As estimated from the model calculations illustrated in Fig.\ \ref{fig_fir} the
typical uncertainty due to the metallicity, IMF slope, and \mup\ is $\sim$ 0.3 dex.

Whereas Eq.\ \ref{eq_fir} should be quite appropriate for starbursts with ages
$\lsim$ 10$^8$ yr, the relation will be more complicated in normal star forming galaxies:
a contribution from old stars to dust heating will lower the coefficient in Eq.\ \ref{eq_fir},
whereas the lower dust optical depth will increase this value.
In this case one may refer to indirect calibrations (``cross calibrations''). E.g.\
for galaxies of Sb types and later one obtains
$ {\rm SFR} = 1.1^{+1.2}_{-0.4} \times 10^{-43} L_{\rm FIR}$ (same units as Eq.\ \ref{eq_fir})
from the work of Buat \& Xu (1996) based on a comparison of IRAS and UV flux 
measurement and appropriate extinction corrections, after a consistent
UV calibration (Eq.\ \ref{eq_uv}) is applied.
In a similar vein Roussel \etal\ (these proceedings) provide a SFR calibration
for measurements using IR filters aboard ISO based on a cross calibration with
an \halpha\ calibration. Their spatial analysis in particular also allows one
to distinguish contributions from old stars and/or AGN in the central regions.

{\em Summary:}
The most important assumptions entering SFR(IR) calibrations are:
the SF history or ``mean age'' of the population and a large dust optical depth.
The ``typical uncertainty'' due to metallicity, IMF slope and \mup\  is $\sim$ 0.3 dex.
One of the main advantages of this method is obviously the negligible effect
of dust extinction.

\subsection{SFR from radio measurements}
The existence of a tight correlation between the radio (1.49 GHz) and FIR luminosity
for normal galaxies has in particular motivated studies of the star formation
rate from radio observations (see the review of Condon 1992).
Possible AGN contamination must be accounted for.
Based on the observed relationship between the radio luminosity and the supernova rate
(cf.\ Condon \& Yin 1990), Condon (1992) has derived a SFR-$L_{\rm radio}$ calibration.
See e.g.\ Cram \etal\ (1998) for a recent discussion of uncertainties related to this 
calibration. 
Synthesis models including radio emission are e.g.\ Mas-Hesse \& Kunth (1981)
and Lisenfeld \etal\ (1996).

The above studies include in general non-thermal and thermal radio emission,
the former dominating in most cases (cf.\ Condon 1992). From a measurement of the
thermal component (if possible) the ionizing photon flux can be derived in a fairly
straightforward way. Its relation to the SFR is discussed in Sect.\ \ref{s_ha}.

Cram \etal\ (1998) present an interesting comparison of SFR indicators from FIR, radio, 
\halpha, and the U-band based on fairly large sample of objects.
Recent examples of applications of observations to studies of the star formation
history of objects up to $z \lsim 1$ are e.g.\ found in Serjeant \etal\ (1998) and
Mobasher \etal\ (1999).

\section{SFR from recombination lines}
\label{s_ha}
Nebular lines re-emit effectively the radiation emitted shortward of the Lyman
limit, and are hence a direct probe of the massive star population. In particular
the hydrogen recombination lines measure directly the total ionizing photon
flux. E.g.\ for Case B recombination the \halpha\ luminosity is given by 
$L($\halpha)$ = 1.36 10^{-12} f_\gamma\, Q_0$, where $f_\gamma$ is the fraction of
photons absorbed by gas and $Q_0$ the total number of Lyman continuum photons ([$s^{-1}$]).
This expression depends only weakly on electron temperature and density
($T_e=$10000 K assumed here). Corresponding expressions for other H recombination
lines are readily derived.

Again, numerous calibrations based on synthesis models are found;
e.g.\  Kennicutt (1983, 1998ab) Gallagher \etal\ (1984), Leitherer \& Heckman (1995),
Gallego \etal\ (1996), Madau \etal\ (1998).
According to K98 the calibrations are typically within $\lsim$ 30 \% when placed on the
same IMF scale and assuming SF equilibrium (see below).
Differences reflect usually changes in stellar and atmosphere models.
An exception is the calibration by Alonso-Herrero \etal\ (1996), used e.g.\ in the study
of Guzman \etal\ (1998), which yields SFRs lower by a factor 2.5. 
The difference appears to be due to difficulties with the former version of the Bruzual 
\& Charlot (1993) models used by these authors (Alonso-Herrero 1999, private communication);
although its origin is not yet clear, 
this difference is not present in their latest model version (cf.\ Madau \etal\ 1998).

\begin{figure}[tb]        
\centerline{
\psfig{figure=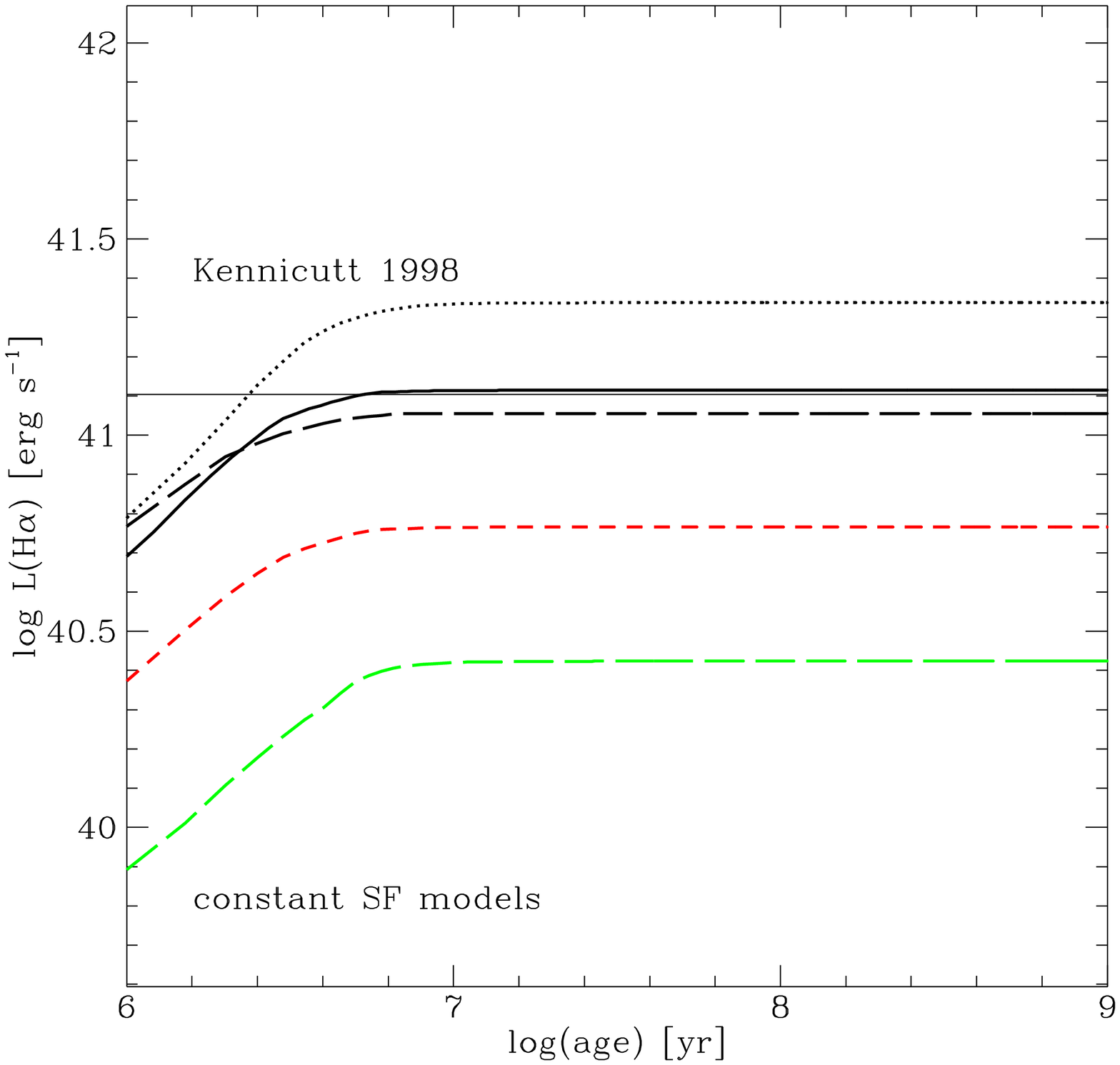,width=8cm}
\psfig{figure=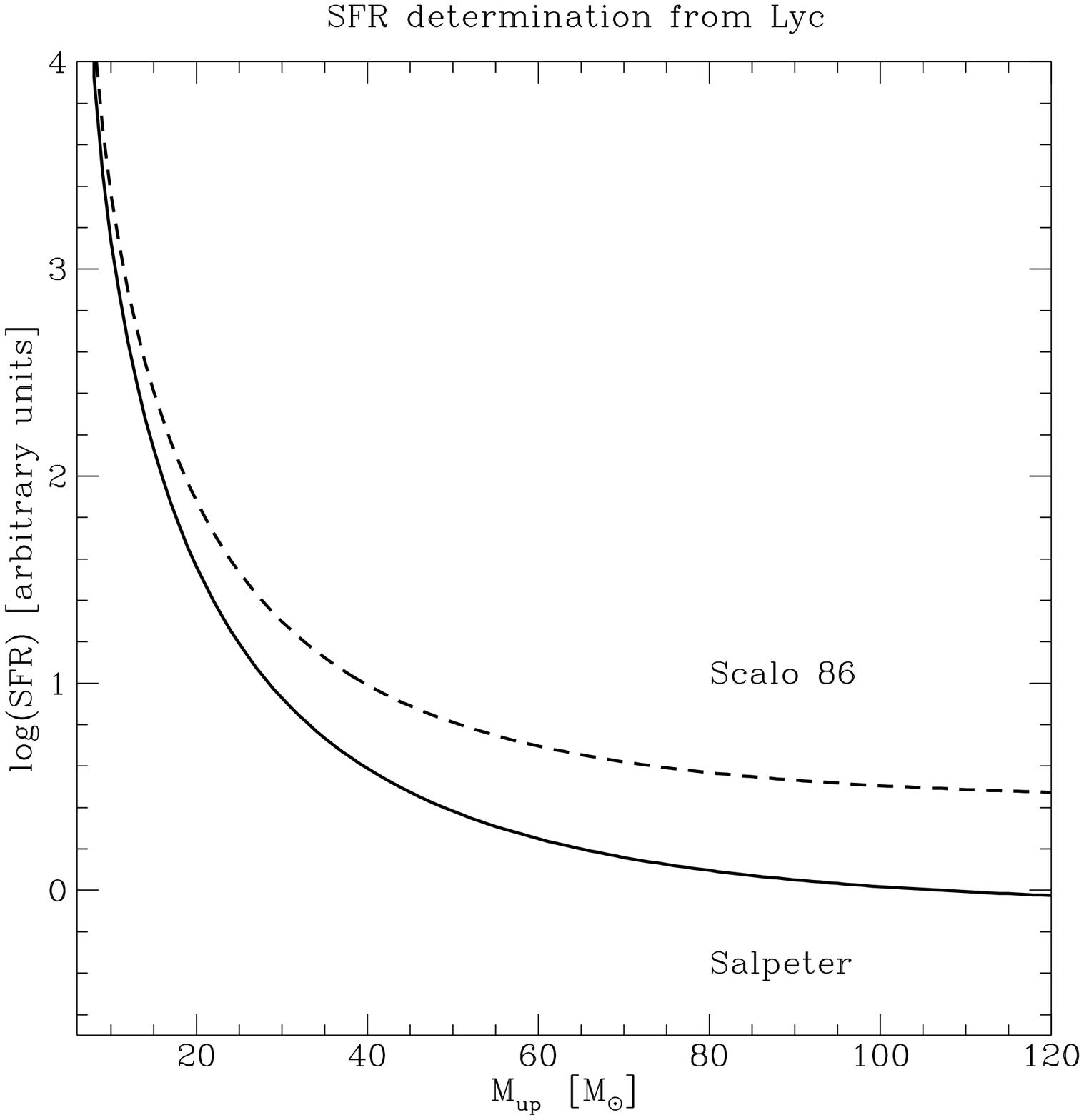,width=8cm}}
\caption{{\em Left:} same as Fig.\ \ref{fig_fir} for \halpha. 
{\em Right:} Variation of the SFR(\halpha) calibration (in arbitrary units) with the 
upper mass cut-off for a Salpeter and Scalo (1986) IMF.}
\label{fig_ha}
\end{figure}

Again using the same IMF as earlier, K98 obtains the following
calibration:
\begin{equation}
  \label{eq_ha}
  {\rm SFR} (\msun \, {\rm yr}^{-1}) = \frac{7.9 \times 10^{-42}}{f_\gamma}
            \frac{\lhalpha}{\rm erg \, s^{-1}} \, A(\halpha)
            =  \frac{1.08 \times 10^{-54}}{f_\gamma} \, Q_0  \, A(H\alpha)
\end{equation}
where $A(\halpha)$ stands for the extinction at \halpha.
This value is derived from a constant SF model at ``equilibrium'', which is 
reached on a very short timescale ($\lsim$ 10 Myr, see Fig.\ \ref{fig_ha}) 
due to the short lifetime of massive stars ($\gsim$ 20 \msun) responsible 
for the ionizing flux.
This relation may also be valid for an ensemble of burst populations
of finite duration, provided enough populations of all ages up to $t_{\rm eq}$
are sampled.
As shown in Fig.\ \ref{fig_ha} variations by typically a factor 2 are obtained
for metallicities $Z$ between 1/20 and 2 \zsun. E.g.\ at lower $Z$ a given population
produces more ionizing flux, i.e.\ the SFR should be corrected downwards.
Note that this difference is predominantly due to changes in the evolutionary tracks
which are predicted to be hotter on average; changes in atmosphere models are
of minor importance (e.g.\ Mas-Hesse \& Kunth 1991).

Due to the strong increase of the ionizing luminosity with stellar mass which largely
overwhelms the lifetime decrease, SFR indicators based on a measure of the Lyman
continuum flux depend strongly on \mup\ and the slope of the upper end of the IMF.
This fact which is rarely appreciated, is illustrated in both panels of Fig.\ 
\ref{fig_ha} (see also Leitherer 1990).
In particular the right panel 
shows the sensitivity of the \halpha\ indicator on \mup\ for a Salpeter and Scalo (1986)
IMF\footnote{The analytic fits of Schaerer (1998) for the ionizing luminosity and
lifetime have been used for this plot.}. 
E.g.\ a decrease e.g.\ \mup\ from 100 to 60 \msun\ implies an increase of the SFR
by a factor of $\sim$ 2 for a Salpeter IMF.

Extinction taken apart, other potential difficulties affecting the SFR(\halpha)
indicator are the possible escape of ionizing photons from the 
observed region/galaxy and the existence of dust {\em inside} the \hii\ regions.
From studies of individual \hii\ regions and diffuse ionized gas in nearby galaxies
(e.g.\ Oey \& Kennicutt 1997, Ferguson \etal\ 1996, and references in K98) escape 
fractions of up to $\sim$ 15--50 \% are found.
Regarding the escape from the entire galaxy this value is probably lower
(e.g.\ Leitherer \etal\ 1995, Deharveng \etal\ 1997).
The quantity of dust inside \hii\ regions competing for the absorption of ionizing 
photons is not well known. Often quoted value are $\sim$ 25 \% from Smith \etal\
(1975). New results from ISO observations should hopefully become available in the
near future.  

{\em Summary:}
The most important assumptions entering SFR(\halpha) calibrations are:
the upper mass cut-off and the slope of the upper end of the IMF.
This method has the shortest equilibrium timescale ($\lsim$ 10 Myr) and is 
therefore independent on the SF history except for very ``local'' applications.
The ``typical uncertainty'' due to metallicity, IMF slope, and \mup\  is $\sim$ 0.7 dex.

\section{SFR from forbidden lines}
\label{s_oii}

In contrast to recombination lines, the emission of forbidden and fine-structure 
lines is not directly coupled to the ionizing luminosity, but depends on
the ionization parameter (reflecting a combination of the ionizing luminosity,
gas density, filling factor, and geometry) and the chemical composition of the
gas. 
Therefore one usually resorts to empirical calibrations.
The strong \oii\ forbidden-line doublet discussed next is frequently used,
in particular since it is accessible to optical observations over a wide range of 
redshifts.
The potential use of other lines, including IR fine-structure lines, and
results from theoretical calibrations are presented in Sect.\ \ref{s_new}.

\subsection{\Oii}
Three empirical ``cross'' calibrations of the \oii\ indicator through \halpha\
are found in the literature. 
Gallagher \etal\ (1989) find the empirical relation $f(OII)=3.2 f(H\beta)$ 
from a sample of irregular galaxies. 
For a sample of normal and peculiar galaxies Kennicutt (1992) obtains: 
\Oii/([N~{\sc ii}]+\halpha)=0.31 and [N~{\sc ii}]/\halpha=0.5.
From the Terlevich \etal\ (1991) sample of \hii\ galaxies Guzm\'an \etal\ (1997)
find: $\langle \lhalpha A(\halpha)/\loii \rangle= 7.42 \pm 1.23$.
The first two relations represent the mean values uncorrected for extinction;
the last expression includes individual extinction corrections.
Adopting for consistency a common \halpha\ calibration 
(Eq.\ \ref{eq_ha} with $f_\gamma=1$) one obtains:
\begin{eqnarray}
  \label{eq_oii}
  {\rm SFR} (\msun \, {\rm yr}^{-1}) & = & 8.9 \times 10^{-42} \,
             \frac{L(OII)}{{\rm erg \,s^{-1}}} \, A(H\alpha) \label{eq_gal} 
             \\ 
      & = & 1.7 \times 10^{-41} \,
  \frac{L(OII)}{{\rm erg \,s^{-1}}} \, A(H\alpha) 
          = 4.7 \times 10^{-41} \, \frac{L(OII)}{{\rm erg \,s^{-1}}} \label{eq_k92}\\
      & = & 5.9 \times 10^{-41} \, \frac{L(OII)}{{\rm erg \,s^{-1}}} \label{eq_guz}
\end{eqnarray}
for the Gallagher \etal, Kennicutt (1992) and Terlevich \etal\ samples
respectively. K98 suggests the use of an average of Eqs.\ \ref{eq_gal} and \ref{eq_k92}
with a coefficient $c=1.4 \pm 0.4 \times 10^{-41}$.
As an example Eq.\ \ref{eq_k92} is also given applying the average 
extinction correction of 1.1 mag derived for nearby spirals (Kennicutt 1983, 1992).
Note, however, that such a correction is only consistent if derived from the
entire sample used to establish the empirical \Oii-\halpha\ relation.

What is the origin of the differences between the above calibrations ?
The higher SFR obtained for the Terlevich \etal\ (1991) sample could be
due to a breakdown of the SF equilibrium assumption entering Eq.\ \ref{eq_ha}.
Indeed it is well known that these objects are mostly of bursting nature as 
opposed to long lasting star formation (e.g.\ Stasi\'nska \& Leitherer 1996).
This fact and a bias for preferentially young objects could 
explain why a larger SFR is obtained.
Other possible explanations are mentioned below.


Some of the uncertainties of the \Oii\ SFR indicator are discussed in K98.
First the calibrated properties (\oii/\halpha, [N~{\sc ii}]/\halpha) show a considerable
scatter in the samples used. Second, \Oii\ may be enhanced by contribution from the
diffuse ionized gas in starburst galaxies.
More generally, however, one must make sure that the samples used for the above
calibration are indeed representative for the considered application.
From a comparison of \luv\ and \loii, Cowie \etal\ (1997) find a reasonable agreement
with the Gallagher \etal\ calibration including a modest extinction correction.
Indirect evidence that the Kennicutt (1992) relation (Eq.\ \ref{eq_k92}) including
$A(\halpha)=$ 1.1 mag may not apply to
the CFRS sample (redshifts 0 $\le z \le$ 1.3) has been found by Hammer \etal\ (1997), 
although the excess of the present stellar mass density found by these authors 
could also simply be due to the neglect of the flattening of the IMF at low masses 
(see Sect.\ \ref{s_gen}).
Clear evidence for considerable variations of the relation between \Oii\ and
\halpha\ emission is found by Jansen (these proceedings) in their 
``Nearby Field Galaxy Survey'' (Jansen \etal\ 1999).
The properties of galaxies from the Stromlo-APM survey are discussed 
by Loveday (these proceedings) and Tresse \etal\ (1999).

As K98 we conclude that \oii\ provides a very useful estimate of the SFR in distant 
galaxies, and is especially useful for a consistency check on other SFR indicators. 
With the help of more complete samples, systematics of the \Oii\ indicator will be 
better understood and the accuracy of calibrations improved. 
New calibrations, based e.g.\ on correlations with \luv\ (e.g.\ Cowie \etal\ 1997,
Hammer \& Flores 1997, Fig.\ 1) might also be used.
Alternatively, new insight can be gained from theoretical modeling.
Such an approach is presented below.

\begin{figure}[tb]        
\centerline{
\psfig{figure=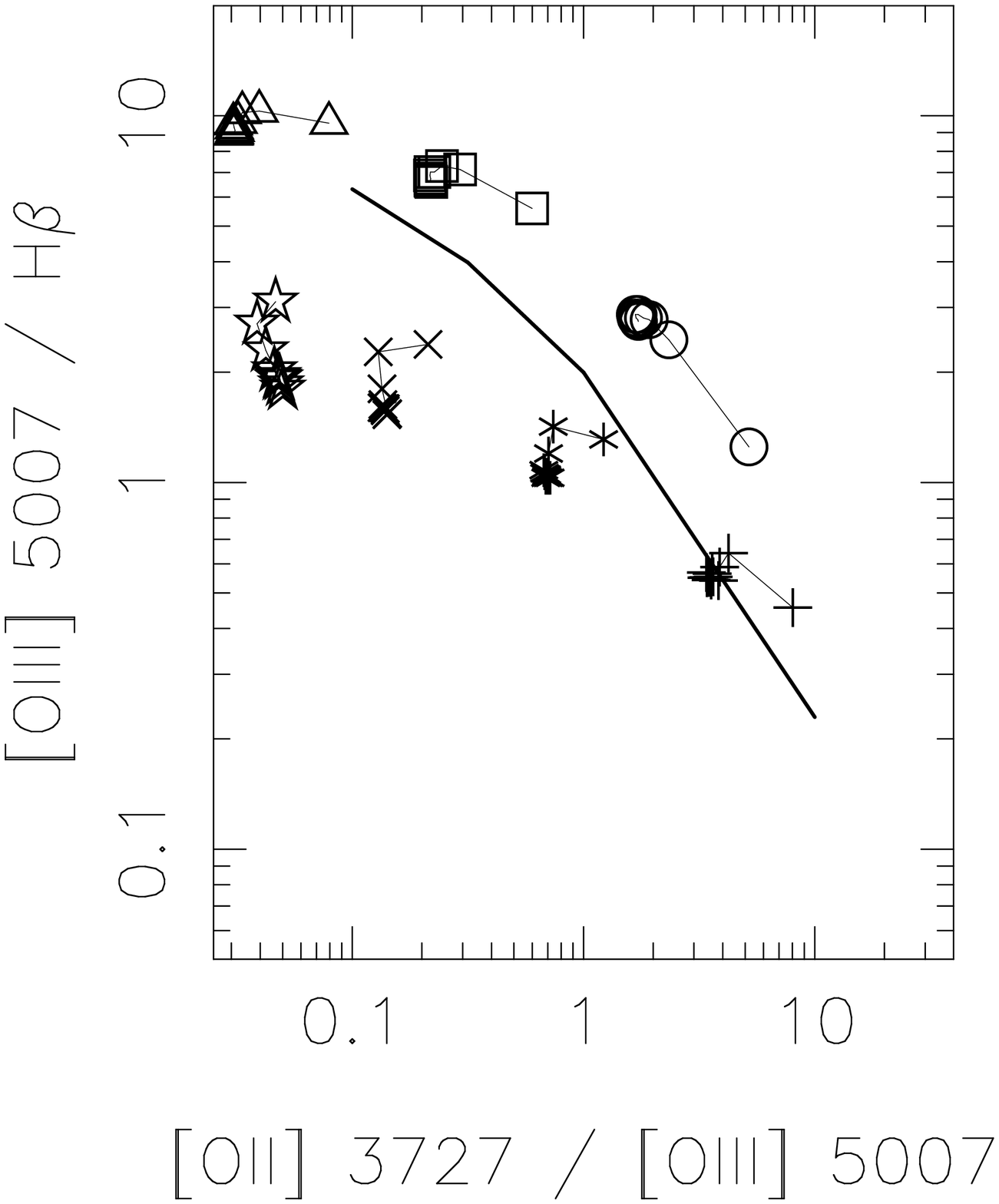,width=8cm,height=8cm}
\psfig{figure=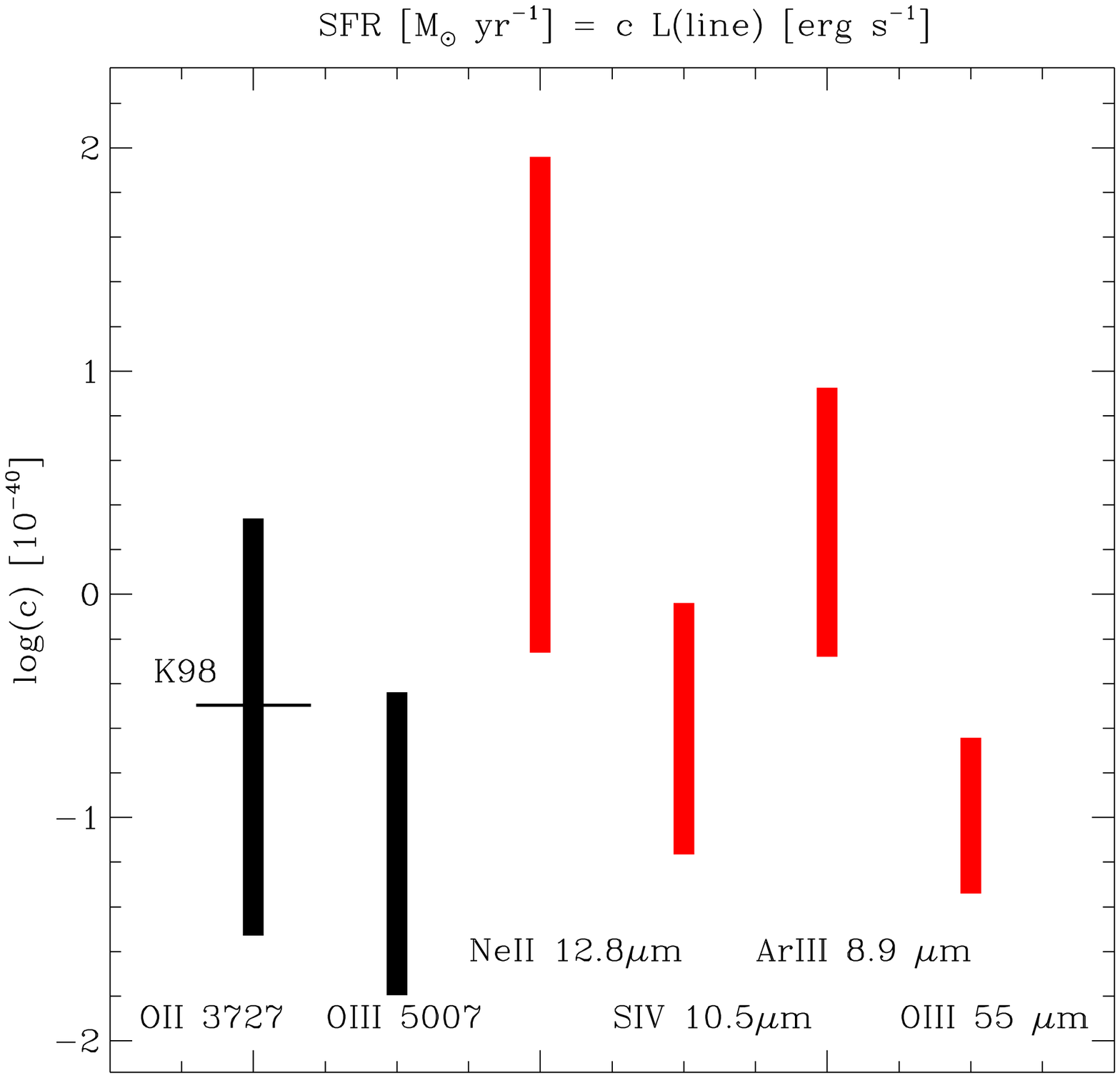,width=8cm}}
\caption{{\em Left:} Emission line ratio 
diagram for representative models with $Z=$0.004 (circles, 
squares, triangles) and 0.02 (solar; plus, asterisk, cross, star) and increasing 
ionization parameter. The solid line indicates the sequence of Baldwin \etal\ (1981).
{\em Right:} Range of $\log c$ (in units of $10^{-40}$, for a 0.1-100 \msun\ Salpeter 
IMF) predicted for various forbidden optical and IR fine-structure lines from models 
in the left panel. The empirical \Oii\ calibration (excluding extinction) of K98 is 
also indicated.}
\label{fig_neb}
\end{figure}

\section{A new approach to forbidden and fine-structure lines}
\label{s_new}
To study the behaviour of \oii\ and other potential SFR indicators from
optical and IR lines we have calculated extensive grids of photoionization
models for starbursts covering different SF scenarios, various metallicities,
and variations of the ionization parameter, gas density, and nebular geometry.
Ionizing fluxes are calculated using the synthesis models of Schaerer \& Vacca 
(1998). We here present preliminary results (see Schaerer \& Stasi\'nska, 
in preparation; cf.\ Charlot 1998 for a similar approach).

Fig.\ \ref{fig_neb} (left) illustrates an emission line diagram for 
representative models with constant SF, metallicities $Z=$0.004 and 0.02 (solar),
and varying ionization parameter.
As expected theoretically, no simple dependence of \loii\ on SFR is found.
For $SFR = c \, \loii$, $c$ is primarily a strong function of metallicity and 
the ionization.
For a given metallicity $c$ increases towards models of higher excitation,
reflecting the decreasing fraction of nebular luminosity emitted by the
\oii\ line.
Over the parameter space explored by the models of Fig.\ \ref{fig_neb} the
``calibration constant'' $c$ varies by a factor of up to 70 (cf.\ right panel)!
This covers probably a parameter space larger than that populated by ``real objects''.
For a more realistic estimate of the uncertainty of SFR(\Oii) or theoretical
calibrations various observational constraints must be taken into account.

Many other nebular lines could in principle serve as SFR indicators
(cf.\ Sodr\'e \& Stasi\'nska 1999). 
In particular the use of IR fine structure lines probing the youngest stellar 
populations could be of interest 
for objects where the IR hydrogen recombination lines cannot be measured
or  as an alternative or complement to IR 
continuum methods, whose calibration depends strongly on the SF history
(Sect.\ \ref{s_ir}).
As for \oii, the expected range for $c$ for some of the strongest metal lines 
in the optical and IR 
([O~{\sc iii}] $\lambda$5007, [Ar~{\sc iii}] 8.9 \micron, [Ne~{\sc ii}] 12.8 \micron, 
[S~{\sc iv}] 10.5 \micron, [O~{\sc iii}] 52 \micron) are shown in Fig.\ \ref{fig_neb} (right).
Work is underway to understand the behaviour of these potential indicators.
%
In addition to the complexity of the parameter space of these models which must be
taken into account for such ``calibrations'', the IR lines are in particular affected
by uncertainties related to atomic data of Ne, Ar, and S (cf.\ Oliva \etal\ 1996,
Schaerer \& Stasi\'nska 1999).
More detailed work on the modeling of IR lines is urgently needed to allow reliable
quantitative studies of the massive star content from IR spectra
(cf.\ Lutz \etal\ 1998, Schaerer \& Stasi\'nska 1998).
%

\section{Summary and conclusions}
\label{s_con}
We have reviewed the major SFR indicators (UV, FIR, \halpha, \oii) as well as the 
methods, assumptions, and uncertainties underlying these tools.
All indicators rely on a calibration with evolutionary synthesis models, which
depend in turn on assumptions on the IMF, the SF history, metallicity $Z$, stellar tracks
and atmospheres.
Extinction, which in particular strongly affects UV studies, is not discussed here.
To quantify a ``typical uncertainty'' related to assumptions on $Z$, the IMF slope and \mup\
test calculations have been presented (parameters given in Table \ref{ta_pars}).
Uncertainties defined in this way are $\sim$ 0.3 dex or larger for all cases.

The FIR method show the strongest dependence on the assumed SF history.
Measurements related to UV light do so to a lesser degree, since the associated timescales 
are shorter ($\sim 10^{8-9}$ yr). 
The \halpha\ indicator (and in principle also \Oii) measuring the ionizing continuum 
produced by massive stars
provide the best measure of instantaneous ($\lsim$ 10 Myr) SF. However, the tradeoff of these
methods is a much stronger dependence on the upper end of the IMF (slope, \mup), which
{\em per se} is only a bad tracer of the total mass, mostly locked up in low mass stars.

Empirical ``cross'' calibrations of SFR indicators from the IR, radio, and the forbidden 
\oii\ lines have also been summarised. For the latter we have in particular discussed
the dependence on the calibration sample.
Using theoretical starburst and photoionization models we have illustrated 
expected variations in calibrations of forbidden lines and pointed out the interest
of IR fine-structure lines for SFR determinations.

The accuracy of SFR determinations can obviously be improved by the use of 
multi-wavelength observations which allow to constrain and test at least some of the 
assumptions made in the calibrations.
Sound comparisons should consider different SF indicators for the same object.
Work in this direction has begun (e.g.\ Meurer \etal\ 1995, Cram \etal\ 
1998, Pettini \etal\ 1998, Glazebrook \etal\ 1999, and references in Sect.\ \ref{s_oii}).
Although {\em relative} comparisons of the SFR in different environments, at different 
redshifts etc.\ can reasonably be made at present times, the slope and cut-off of the 
IMF at low masses remains the major uncertainty in determinations of {\em absolute} star 
formation rates.
Both theoretical and observational progress should allow to improve our knowledge
on the IMF and possible variations of it, and to study the process of star formation from
the distant to the local universe.

\medskip
\noindent
{\small {\sl Acknowledgments}
Jonathan Braine, Claus Leitherer, Gerhard Meurer and Grazyna Stasinska provided 
useful comments on an earlier version of the manuscript.
}

\vfill

{\small
\begin{moriondbib}
\bibitem{} Alonso-Herrero, A., et al., 1996, \mnras {278}{417}
\bibitem{} Baldwin, J., Philips, M., Terlevich, R., 1981, {\sl PASP} {\bf 93}, 5
\bibitem{} Bruzual, G., Charlot, S., 1993, \apj {405}{38}
\bibitem{} Bruzual, G., Charlot, S., 1998, in preparation
\bibitem{} Buat, V., 1992, \aa {264}{444}
\bibitem{} Buat, V., Burgarella, D., 1998, \aa {334}{772}
\bibitem{} Buat, V., Deharveng, J.M., Donas, J., 1989, \aa {223}{42}
\bibitem{} Buat, V., Xu, C., 1996, \aa {306}{61}
\bibitem{} Calzetti, D., Kinney, A.L., Storchi-Bergmann, T., 1994, \apj {429}{582} 
\bibitem{} Calzetti, D., Kinney, A.L., Storchi-Bergmann, T., 1996, \apj {458}{132} 
\bibitem{} Charlot, S., 1998, in {\sl NGST - Science Drivers \& Technonogical Challenges}, 
        34th Liege Astrophysics Colloquium, eds. Benvenuti, P. et al., ESA-SP 429, p. 135
\bibitem{} Condon, J.J., 1992, {\sl ARA\&A} {\bf 30}, 575
\bibitem{} Cram, L., et al., 1998, \apj {507}{155}
\bibitem{} Cowie, L.L, et al., 1997, \apj {481}{L9}
\bibitem{} Deharveng, J.M., et al., 1994, \aa {289}{715}
\bibitem{} Deharveng, J.M., et al., 1997, \aa {325}{1259}
\bibitem{} Dorman, B., O'Connell, R.W., Rood, R.T., 1995, \apj {442}{105}
\bibitem{} Ferguson, A.M.N., et al., 1996, \aj {111}{2265}
\bibitem{} Gallagher, J.S., Bushouse, H., Hunter, D.A., 1989, \aj {97}{700}
\bibitem{} Gallagher, J.S., Hunter, D.A., Tutukov, A.V., 1984, \apj {284}{544}
\bibitem{} Gallego, J., et al., 1995, \apj {455}{L1}
\bibitem{} Gilmore, G., Parry, I., Ryan, S., (Eds.), 1998, {\sl ASP Conf. Series} {\bf 142}
\bibitem{} Glazebrook, K., et al., 1999,\mnras {in press}{(astro-ph/9808276)}
\bibitem{} Goldader, J.D., et al., 1997, \apj {474}{104}
\bibitem{} Guzm\'an, R., et al., 1997, \apj {489}{559}
\bibitem{} Hammer, F., Flores, H., 1998, in {\sl Dwarf Galaxies and Cosmology}, XVIIIth 
  Moriond astrophysics meeting, in press, (astro-ph/9806184)
\bibitem{} Hammer, F., et al., 1997, \apj {481}{49}
\bibitem{} Hunter, D.A., et al., 1986, \apj {303}{171} 
\bibitem{} Jansen, R.A., et al., 
  1999, \apjs {submitted}
\bibitem{} Kennicutt, R.C., 1983, \apj {272}{54}
\bibitem{} Kennicutt, R.C., 1992, \apj {388}{310}
\bibitem{} Kennicutt, R.C., 1998a, {\sl ARA\&A} {\bf 36}, 189 (K98)
\bibitem{} Kennicutt, R.C., 1998b, \apj {498}{541}
\bibitem{} Kroupa, P., 1998, {\sl ASP Conf. Series} {\bf 134}, 483
\bibitem{} Kroupa, P., et al., 1993, \mnras {262}{545}
\bibitem{} Leitherer, C., 1990, \apjs {73}{1}
\bibitem{} Leitherer, C., Heckman, T.M., 1995, \apjs {96}{9}
\bibitem{} Leitherer, C., et al., 1997, \apj {454}{L19}
\bibitem{} Leitherer, C., et al., 1999, \apjs {in press} {(astro-ph/9902334)}
\bibitem{} Lisenfeld, U., V\"olk, H.J., Xu, C., 1996, \aa {314}{745}
\bibitem{} Lutz, D., et al., 1998, {\sl ASP Conf. Series} {\bf 132}, 89
\bibitem{} Madau, P., Pozetti, L., Dickinson, M., 1998, \apj {498}{106}
\bibitem{} Mas-Hesse, J.M., Kunth, D., 1991, \aas {88}{399} 
\bibitem{} Mateo, M.L., 1998, {\sl ARA\&A} {\bf 36}, 435
\bibitem{} Meurer, G., et al., 1995, \aj {110}{2665}
\bibitem{} Meurer, G., et al., 1997, \aj {114}{54}
\bibitem{} Meurer, G., et al., 1999, \apj {in press}{(astro-ph/9903054)}
\bibitem{} Mobasher, B., et al., 1999, \mnras {in press}{(astro-ph/9903293)} 
\bibitem{} Oey, M.S., Kennicutt, R.C., 1997, \mnras {291}{827}
\bibitem{} Oliva, E., Pasquali, A., Reconditi, M., 1996, \aa {305}{L21}
\bibitem{} Pettini, M., et al., 1998, \apj {508}{539}
\bibitem{} Reid, I.N., Gizis, J., 1997, \aj {113}{2246}
\bibitem{} Rieke, G.H., et al., 1980, \apj {238}{24}
\bibitem{} Scalo, J., 1986, {\sl Fund. Cosm. Phys.} {\bf 11}, 1 
\bibitem{} Scalo, J., 1998, {\sl ASP Conf. Series} {\bf 142}, 201
\bibitem{} Schaerer, D., 1998, {\sl ASP Conf. Series} {\bf 131}, 310
\bibitem{} Schaerer, D., Stasi\'nska, G., 1998, in {\sl The Universe as seen by ISO}, 
  Eds. P. Cox, M.F. Kessler, ESA Special Publications series (SP-427),{\bf  in press}, 
  (astro-ph/9812068)
\bibitem{} Schaerer, D., Stasi\'nska, G., 1999, \aa {345}{L17}
\bibitem{} Schaerer, D., Vacca, W. D., 1998, \apj {497} {618}
\bibitem{} Serjeant, S., Carlotta, C., Oliver, S., 1998, \mnras {in press}{(astro-ph/9808259)}
\bibitem{} Smith, L.F., Biermann, P., Mezger, P.G., 1978, \aa {66}{65}
\bibitem{} Sodr\'e Jr., L., Stasi\'nska, G., 1999, \aa {in press}{(astro-ph/9903130)}
\bibitem{} Stasi\'nska, G., Leitherer. C., 1996, \apjs {107}{66}
\bibitem{} Terlevich, R., 1991, \aas {91}{285}
\bibitem{} Tresse, L., et al., 1999, \mnras {in press}{(astro-ph/9905384)}
\bibitem{} Thronson, H.A., Telesco, C.M., 1986, \apj {311}{98}

 \end{moriondbib}
}
\end{document}